\documentclass[showpacs,twocolumn]{revtex4}
\usepackage[dvips]{epsfig}

\newcommand{\be}{\begin{equation}}
\newcommand{\ee}{\end{equation}}
\newcommand{\clF}{{\cal F}}

\newcommand{\clL}{{\cal L}}
\newcommand{\clP}{{\cal P}}

\newcommand{\bea}{\begin{eqnarray}}
\newcommand{\eea}{\end{eqnarray}}

\newcommand{\prt}{\partial}
\newcommand{\tlP}{\tilde{P}}
\newcommand{\rgl}{\rangle}
\newcommand{\lgl}{\langle}

\newcommand{\hV}{\hat{ L}_{FP}}

\begin{document}

\title{Subdiffusion on a Fractal Comb}
\author{Alexander Iomin}

\affiliation{Department of Physics, Technion, Haifa, 32000,
Israel}

\date{[Phys. Rev. E \textbf{83}, 052106 (2011)]}
\begin{abstract}
Subdiffusion on a fractal comb is considered. A mechanism of
subdiffusion with a transport exponent different from $1/2$ is
suggested. It is shown that the transport exponent is determined
by the fractal geometry of the comb.
\end{abstract}


\maketitle

A comb model was introduced for understanding of anomalous
transport in percolating clusters \cite{arch9,em1} and it was
considered as a toy model for a porous medium used for exploration
of low dimensional percolation clusters \cite{arch9,baskin1}, as
well. It is a particular example of a non-Markovian phenomenon,
which was explained in the framework of continuous time random
walks \cite{em1,shlesinger,klafter,dubkov,sokolov}. In the last
decade the comb model has been extensively studied for
understanding of different realizations of non-Markovian random
walks both continuous \cite{Arkhincheev,daSilva,bi2004} and
discrete \cite{Cassi}.

Usually, anomalous diffusion on the comb is described by the $2D$
distribution function $P=P(x,y,t)$, and a special behavior is that
the displacement in the $x$--direction is possible only along the
structure axis ($x$-axis at $y=0$). Therefore, diffusion in the
$x$-direction is highly inhomogeneous. Namely, the diffusion
coefficient is $D_{xx}=\tilde{D}\delta(y)$, while the diffusion
coefficient in the $y$--direction (along fingers) is a constant
$D_{yy}=D$. Therefore, this inhomogeneous diffusion is described
by the Fokker-Planck equation in the dimensionless time and
coordinates
\be\label{FC_a}  %
\hV P(x,y,t)\equiv\prt_t P-\delta(y)\prt_x^2 P -\prt_y^2P=0 \, . \ee %
It is obtained by the rescaling with relevant combinations of the
comb parameters $D$ and $\tilde{D}$, such that the dimensionless
time and coordinates are $D^3t/\tilde{D}^2\rightarrow t$
$Dx/\tilde{D}\rightarrow x$, $Dy/\tilde{D}\rightarrow y$,
correspondingly \cite{ib2005}.

The fractional transport along the structure $x$ axis is described
by the transporting contaminant distribution
$p(x,t)=\int_{-\infty}^{\infty}P(x,y,t)dy$. It was shown
\cite{ib2005} that Eq. (\ref{FC_a}) is equivalent to the
fractional Fokker-Planck equation
\be\label{FC_b} %
\prt_t^{\frac{1}{2}}p(x,t)-\frac{1}{2}\prt_x^2 p(x,t)=0 \, , %
\ee  %
from where subdiffusion can be immediately obtained: $\int
x^2p(x,t)dx\sim \sqrt{t}$. Here $\prt_t^{\frac{1}{2}}$ is a
fractional time derivative, which is a formal notation of an
integral with a power law memory kernel. For $0<\alpha<1$ it reads
\begin{equation}\label{fse2}
\prt_t^{\alpha}f(t)
=\int_0^t\frac{(t-\tau)^{-\alpha-1}}{\Gamma(1-\alpha)}
\prt_{\tau}f(\tau)d\tau\, . %
\end{equation} %
Subdiffusive mechanism with an arbitrary transport exponent was
also suggested by changing either the boundary conditions for
diffusion in the fingers
\cite{benAvraham,Chukbar,Dvoretskaya,Reynolds}, or introducing a
dependence of the diffusion coefficient on time and space
\cite{Zahran}. In this paper we consider a fractal comb, when
diffusion is highly inhomogeneous along the $y$ fingers, as well.
Namely, it takes palace for those coordinates of the $x$ axis,
which belong to a fractal set $S_{\nu}(x)$, and is defined by a
characteristic function $\chi(x)$, such that $D_{yy}=D\chi(x)$,
where $\chi(x)=1$, if $x\in S_{\nu}(x)$ and $\chi(x)=0$, if
$x\notin S_{\nu}(x)$. The fractal set $S_{\nu}(x)$ is a random
fractal with a fractal dimension $0<\nu<1$ embedded in the $1D$
Euclidian space (of the $x$ axis). Such generalization of the comb
model to a discrete (fractal) comb model for consideration of
fractional transport in discrete systems is more realistic
situation for theoretical studies of transport properties in
discrete systems with complicated topology including fractal ones
like  porous discrete media \cite{baklanov}, electronic transport
in semiconductors with a discrete distribution of traps, cancer
development with definitely fractal structure of the spreading
front , see  \textit{e.g.}, reviews \cite{sokolov,gouyet}, and
infiltration of diffusing particles from one material to another
\cite{korabel}.

Hence, we study the following dimensionless equation
\be\label{FC_1}  %
\prt_t \clP-\delta(y)\prt_x^2\clP-\chi(x)\prt_y^2\clP=0 \, .
\ee %
The initial condition is $ \clP(x,y,0)=\delta(x)\delta(y)$, and
the boundary conditions on infinities have the form
$\clP(\pm\infty,\pm\infty,t)=0$ and the same for the first
derivatives with respect to $x$ and $y$
$\clP_x^{\prime}(\pm\infty,\pm\infty,t)=
\clP_y^{\prime}(\pm\infty,\pm\infty,t) =0$.

Our main purpose is to evaluate the second moment
\be\label{FC_sigma2}  %
\lgl x^2\rgl(t)=\int x^2\clP(x,y,t)dxdy\,  \ee  %
as a function of time. Therefore, the forthcoming analysis of Eq.
(\ref{FC_1}) is supposed to be carried out under the integration
sign. Using properties of the characteristic function
\be\label{FC_2}  %
\chi^2(x)=\chi(x)~~~\mbox{and} ~~~\prt_x\chi(x)=0\, ,
\ee    %
a solution of Eq. (\ref{FC_1}) can be presented in the form
\be\label{FC_3} %
\clP(x,y,t)=\chi(x)P(x,y,t) \, , \ee  %
where $P(x,y,t)$ is a solution of the continuous comb model and we
shall show that an equation for this function coincides with Eq.
(\ref{FC_a}). But first, one should understand a physical meaning
of the distribution $\clP(x,y,t)$ based on properties of the
characteristic function $\chi(x)$. While the first property in Eq.
(\ref{FC_2}) is obvious and follows from the definition of
$\chi(x)$, the second expression deserves an explanation. To show
this, let us consider the $N$th step of the fractal set $S_{\nu}$
construction. It is a union of disjoint intervals $\Delta x_N$. In
general case of a random fractal, these are random intervals. In
the limiting case one obtains
$S_{\nu}=\lim_{N\to\infty}\bigcup\Delta x_N$. Therefore, the
characteristic function on every interval $\Delta
x_j=[x_j,x_j+\Delta x_N]$ is $\chi(\Delta x_N)=
\Theta(x-x_j)-\Theta(x-x_j-\Delta x_N)$. Differentiation of the
characteristic function on every interval yields $\frac{\prt}{\prt
x}\chi(\Delta x_N)=\delta(x-x_j)-\delta(x-x_j-\Delta x_N)$. In the
limit $N\to\infty$ it tends to zero (under the integration), since
$P(x,y,t)$ and its derivatives are continuous functions.

\begin{figure}
\begin{center}
\epsfxsize=6.6cm \leavevmode
    \epsffile{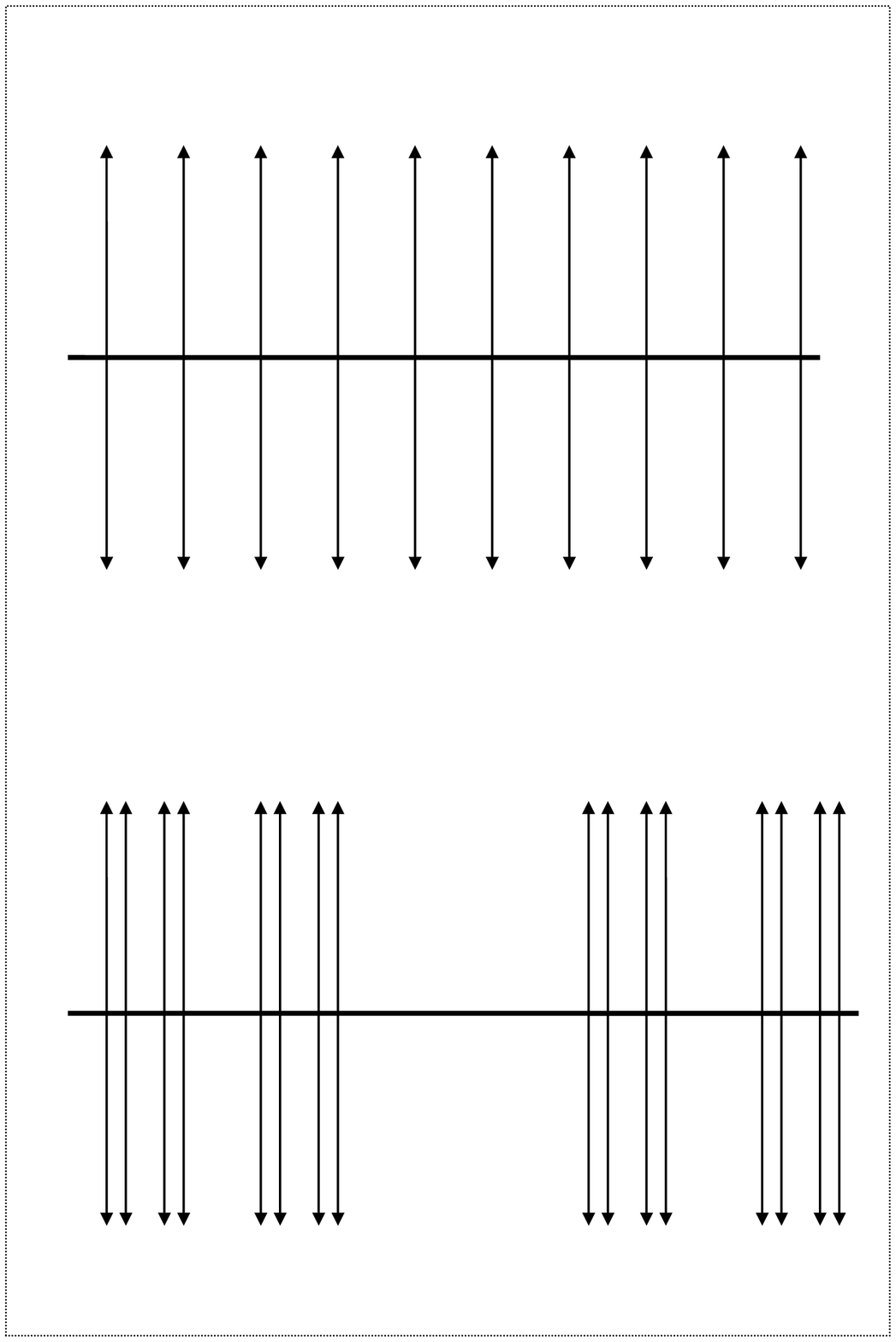}
\caption{Comb structures. The upper strip corresponds to the
continuous comb model. The lower strip is a sketch of the fractal
comb with a specific distribution of fingers, which corresponds to
the one third Cantor set (at the forth step of the construction).
Asking a question: ``one third of what scale on the infinite line
is it?'', one should recognize that this construction is not
representative. Therefore, the fractal set $F_{\nu}$ can be
considered as a random fractal distribution of the fingers without
any specifically defined construction algorithm. It is worth
admitting that the fractal distribution of teeth, in general case,
is multifractal, and the uncertainty of this construction should
be stressed as well. }
\end{center}
\end{figure}

Now we can return to the continuity property of $\clP(x,y,t)$ that
can be understood from the calculation of the second moment $\lgl
x^2\rgl(t)$ in Eq. (\ref{FC_sigma2}). The presence of the
characteristic function in this expression means that the
integration is performed over the fractal volume \cite{tarasov}.
It means that $\int\chi(x)dx\rightarrow\frac{1}{\Gamma(\nu)}\int
|x|^{\nu-1}dx\sim\frac{1}{\Gamma(\nu+1)} |x|^{\nu}$, where
$\Gamma(\nu)$ is the Gamma function. This yields the expression
for $\lgl x^2\rgl(t)$
\be\label{FC_4}  %
\lgl x^2\rgl(t)=\frac{1}{\Gamma(\nu)}
\int_{-\infty}^{\infty}x^2|x|^{\nu-1}P(x,y,t)dxdy\,
.\ee  %
Using this smoothing procedure, we can show that an equation for
the distribution function $P(x,y,t)$ is Eq. (\ref{FC_a}). We have
from Eqs. (\ref{FC_1}) and properties (\ref{FC_2}) of the
characteristic function $\chi(x)$ that for any arbitrary function
$f(x)$
\be\label{FC_5}  %
\int_{-\infty}^{\infty}|x|^{\nu-1}f(x)\hV P(x,y,t)dx=0\, . \ee  %
Therefore, we have equation $\hV P(x,y,t)=0$, which exactly
coincides with Eq. (\ref{FC_a}) and is valid for all $x$.

Now we at a position to determine $\lgl x^2\rgl(t)$. Taking into
account Eq. (\ref{FC_4}), one obtains from integration of Eq.
(\ref{FC_5}) over $y$ with $f(x)=x^2$
\be\label{FC_7}  %
\prt_t\lgl x^2\rgl(t)=\frac{1}{\Gamma(\nu)}
\int_{-\infty}^{\infty}|x|^{\nu-1}x^2\prt_x^2P(x,0,t)dx\,
. \ee %
Here we take into account that
$\int_{-\infty}^{\infty}\prt_y^2P(x,y,t)=0$ due to the boundary
conditions. A relation between $P(x,0,t)$ and $P(x,y,t)$ can be
established in the Laplace domain. Performing the Laplace
transform $\hat{\clL}P(x,y,t)=\tlP(x,y,s)$ in Eq. (\ref{FC_a}), it
is readily to see that $\tlP(x,y,s)=\tlP(x,0,s)e^{-\sqrt{s}|y|}$
satisfies the equation. After integrating over $y$, it yields
\be\label{FC_8} %
 \tlP(x,0,s)=\frac{1}{2}\sqrt{s}\int_{-\infty}^{\infty}\tlP(x,y,s)dy
 =\frac{1}{2}\sqrt{s}\tilde{p}(x,s)\, .\, .
\ee %
This result can be taken into account after the Laplace transform
in Eq. (\ref{FC_7}), that yields an expression for the second
moment in the Laplace domain $\hat{\clL}[\lgl x^2\rgl(t)]=
\widetilde{\lgl x^2\rgl}(s)$. It reads
\be\label{FC_9}  %
s\widetilde{\lgl x^2\rgl}(s)=\frac{1}{\Gamma(\nu)}
\int_{-\infty}^{\infty}|x|^{1+\nu}\prt_x^2\tilde{p}(x,s)dx\, ,
\ee   %
where $\tilde{p}(x,s)=\frac{1}{\sqrt{2s^{3/2}}}
\exp(-\sqrt{2s^{1/2}}|x|)$ can be obtained from Eq. (\ref{FC_b}).
After the Laplace inversion one obtains the second moment
\be\label{FC_10}  %
\lgl x^2\rgl(t)=K_{\nu} t^{\frac{1+\nu}{4}}\, , \ee  %
where $K_{\nu}=\Gamma(2+\nu)/\Gamma(\frac{5}{4}+\frac{\nu}{4})
\Gamma(\nu)\sqrt{2^{1+\nu}}$ is a generalized diffusion
coefficient. Finally, we obtain subdiffusion on the comb $\lgl
x^2\rgl(t)\sim t^{\mu}$ with the transport exponent
$\frac{1}{4}<\mu<\frac{1}{2}$. When $\nu=1$, one observes
subdiffusion with $\mu=1/2$. To obtain subdiffusion with
$\frac{1}{2}<\mu<1$, one considers advection along the structure
$x$ axis instead of diffusion. This yields for the transport
exponent to be $\mu=\frac{1+\nu}{2}$.

The main deficiency of the obtained result in Eq. (\ref{FC_10}) is
that it is based on the presentation of the probability
distribution function as a product of a continuous function and
the characteristic function in Eq. (\ref{FC_3}). Although the
inferring of Eq. (\ref{FC_10}) is correct, this presentation can
leads to wrong result, because the probability distribution
function $\clP(x,y,t)$ must be continuous at every point. To
overcome this deficiency, we refuse the locality property. To this
end, the following procedure of coarse graining of the
Fokker-Planck equation (\ref{FC_1}) is suggested. First, we apply
the Fourier transform to Eq. (\ref{FC_1}) with respect to the $x$
coordinate. To apply this transformation to the last term in Eq.
(\ref{FC_1}), we use the following auxiliary identity
\[\chi(x)f(x)=\prt_x\int_0^x\chi(y)f(y)dy\, , ~~\chi(0)=0\, .\] Here for brevity we
define  $f(x)\equiv \clP(x,y,t)$. This integration with the
characteristic function can be carried out by means of a
convolution \cite{bi2011}. Note that
\be\label{FC_11}  %
\int_0^x\chi(y)f(y)dy=\sum_{x_j\in
S_{\nu}[0,x]}\int_{-\infty}^{\infty}f(y)\delta(y-x_j)dy\, , \ee  %
where we use that \[\sum_{x_j\in
S_{\nu}}\delta(y-x_j)=\mu^{\prime}(x)\sim x^{\nu-1}\] is a fractal
density, such that on the finite interval $(0,x)$, the integral
$\int_0^xd\mu(y)\sim x^{\nu}$ corresponds to the fractal volume.
Therefore, Eq. (\ref{FC_11}) reads
\[\int_0^x\chi(y)f(y)dy=\int_0^xf(y)d\mu(y)\, .\] The last expression
can be rewritten as a convolution integral. Due to Theorem $3.1$
in Ref.~\cite{Ren}, we have
\be\label{FC_12}  %
\int_0^xf(y)d\mu(y)\simeq
\frac{A_{\nu}}{\Gamma(\nu)}\int_0^x(x-y)^{\nu-1}f(y)dy\,
,\ee  %
where $A_{\nu}$ is a constant, defined by the conditions of the
theorem. In sequel we disregard this parameter, putting
$A_{\nu}=1$. This integration is a Riemann-Liouville integral
(see, \textit{e.g.}, \cite{klafter,oldham})
\[\prt_x{}_0I_x^{\nu}f(x)\equiv\frac{1}{\Gamma(\nu)} \prt_x
\int_0^x(x-y)^{\nu-1}f(y)dy \, .\] Here we use a standard notation
${}_0I_x^{\nu}f(x)$ to define integration with a power law kernel,
see Eq. (\ref{fse2}). The integration can be presented in the form
of the inverse Laplace transform $\hat{\clL}f(x)=\tilde{f}(s)$,
which reads \[{}_0I_x^{\nu}f(x)=
\hat{\clL}^{-1}\hat{\clL}[{}_0I_x^{\nu}f(x)]=
\hat{\clL}^{-1}s^{-\nu}\tilde{f}(s)\, .\] Therefore, after the
variable change $s=iz$, the Fourier transform of the last term in
Eq. (\ref{FC_1}) yields
\be\label{FC_13}  %
\hat{\clF}_x[\chi(x)\clP(x,y,t)]=(ik)^{1-\nu}\hat{\clP}(k,y,t)\,
, \ee %
where $\hat{\clF}_xf(x)=\hat{\clP}(k,y,t)=\tilde{f}(ik)$. One
takes into account that the result should be symmetrical with
respect to the negative $x<0$. Therefore, the Fourier transform of
Eq. (\ref{FC_1}) yields
\be\label{FC_14}  %
\prt_t\hat{\clP}(k,y,t)=-\delta(y)k^2\hat{\clP}(k,y,t)+
|k|^{1-\nu}\prt_y^2\hat{\clP}(k,y,t)\, .  \ee  %
Now we perform the Laplace transform with respect to time
$\hat{\clL}\hat{\clP}(k,y,t)=\hat{\tilde{\clP}}(k,y,s)\equiv
G(k,y,s)$. This yields
\be\label{FC_15}  %
sG=-\delta(y)k^2G+|k|^{1-\nu}\prt_y^2G+\delta(y) \ee  %
with the solution
\be\label{FC_16}  %
G(k,y,s)=\exp(-|y|\sqrt{s|k|^{\nu-1}})g(k,s)\, ,\ee   %
where $g(k,s)=G(k,0,s)$ is the Fourier-Laplace image on the
structure axis at $y=0$. Again, we are interesting in dynamics
along the structure axis by studying the probability distribution
function $\clP_1$ (see Eq. (\ref{FC_8})):
\be\label{FC_17}  %
\clP_1(x,t)=\int_{-\infty}^{\infty}\clP(x,y,t)dy\, . \ee %
Integrating Eq. (\ref{FC_16}) over $y$ one obtains
\be\label{FC_17a}  %
G(k,0,s)=\frac{\sqrt{s|k|^{\nu-1}}}{2}
\int_{-\infty}^{\infty}G(k,y,s)dy\, .  \ee   %
Therefore, integrating Eq. (\ref{FC_15}) over $y$ yields the
Montrall-Weiss equation that, after the Fourier and the Laplace
inversions, reduces to the fractional Fokker-Planck equation. It
is a particular case of a general equation
\be\label{FC_18}  %
\prt_t^{\alpha}\clP_1(x,t)=\frac{1}{2}\nabla_x^{\beta}\clP_1(x,t)\,
,~~0\leq\alpha\leq 1\, ,
\ee  %
where $\beta=\frac{3}{2}+\frac{\nu}{2}$. It describes a
competition between long rests and long flights.  We stress that
when $\alpha=1/2$, it corresponds to the comb model, see Eq.
(\ref{FC_b}). Here we use the formal definition for the Riesz-Weyl
fractional space derivative in form of the Fourier inversion (see,
\textit{e.g.}, \cite{klafter,zaslavsky}):
\be\label{FC_18a}  %
\nabla_x^{\nu}f(x)=\hat{\clF}^{-1}[|k|^{\nu}\hat{f}(k)]\, .  \ee  %
This equation was studied in \cite{mk262,mk217} (see also review
\cite{klafter}). The correct form of the mean squared
displacement, which estimates the competition of ``laminar motion
events'' (flights) and ``localization'' (waiting) events in the
L\'evy walk picture was obtained through the relation \cite{mk262}
valid for  Eq. (\ref{FC_18})
\be\label{FC_19}  %
\lgl x^2(t)\rgl\sim t^{\mu}=t^{1-\frac{\nu}{2}}\, .
\ee  %
One easily checks that this result  with $\mu =1-\nu/2$ has a
correct limit for $\nu=1$, when it corresponds to the continuous
comb model with $\mu=1/2$.

In conclusion, we presented two approaches to study subdiffusion
on the fractal comb, when the fractal trap distribution is
determined by the characteristic function $\chi(x)$. In this case,
it is tempting to look for a solution in the multiplicative form
of Eq. (\ref{FC_3}). Thus the forthcoming analysis, based on this
presentation of the probability distribution function is rigorous.
The main deficiency of this approach is that it violates the
continuity property of the probability distribution function. To
overcome this deficiency, a coarse graining procedure of the
Fokker-Planck equation (\ref{FC_1}) is suggested. It is based on
the possibility to perform the Fourier transform for Eq.
(\ref{FC_1}) exactly. We obtained that inhomogeneous, fractal
distribution of traps in the comb model leads to L\'evy jumps that
complicates fractal diffusion and leads to the competition between
long jumps and localization inside traps. This phenomenon is
described by the fractional Fokker-Planck equation (\ref{FC_18}).
As a result of this competition, subdiffusion, which is the
dominant process, realizes.

We admit that the results of the either approach of Eqs.
(\ref{FC_10}) or (\ref{FC_19}) have correct limits for $\nu=1$. A
specific property of Eq. (\ref{FC_19}) is that for $\nu=0$, it
corresponds to normal diffusion with $\mu=1$. While the first
limit with $\nu=1$ is well understood, the second one is not so
obvious. Indeed, when the Hausdorff dimension is $\nu=0$, there
are no traps, and normal diffusion is anticipated. Nevertheless,
this behavior results from the fractional Fokker-Planck equation
(\ref{FC_18}) with $3/2$ fractional space derivative and $1/2$
fractional time derivative. This is a special point of a
transition from subdiffusion to superdiffusion
\cite{klafter,mk262}. For the comb model subdiffusion is the
dominant process, there is no superdiffusion, and $\mu=1$ is the
boundary point.

 This research was supported in part by the Israel Science
Foundation (ISF) and by the US-Israel Binational Science
Foundation (BSF).

\end{document}